\documentclass[copyright]{eptcs}
\providecommand{\event}{T. Neary, D. Woods, A.K. Seda and N. Murphy (Eds.):\\ The Complexity of Simple Programs 2008.}
\providecommand{\volume}{1}
\providecommand{\anno}{2009}
\providecommand{\firstpage}{118}
\usepackage{breakurl}

\usepackage{amssymb}

\title{Some Considerations on Universality}

\author{Manfred Kudlek
    \institute{Dept. Informatik, MIN Fakult\"at, Univ. Hamburg}
    \email{kudlek@informatik.uni-hamburg.de}
}
\def\titlerunning{Some Considerations on Universality}
\def\authorrunning{M. Kudlek}

\begin{document}
\maketitle

\begin{abstract}
The paper puts into discussion the concept of universality, in particular
for structures not of the power of Turing computability. The question arises
if for such structures a universal structure of the same kind exists or not.
For that the construction of universal Turing machines and those with some
constraints are presented in some detail.
\end{abstract}


\section{Introduction}

Without the investigation of the concept of {\it universality} it is quite
possible that our modern computers wouldn't exist. Although they
are finite automata, at least theoretically, practically they can be treated
as Turing machines processing any special program which can be seen as a
special Turing machine. 
This paper presents some general considerations on that concept, presenting
in some more detail universality for general Turing machines, deterministic as well
as non-deterministic ones. Universal devices not only have been considered for
Turing machines (TM's), but also for cyclic Post machines (CPM's), cellular automata
(CA's), random access machines (RAM's), and others.

Very small universal devices mainly have been constructed for DTM's, CPM's, and RAM's.
Until recently, the encoding of special deterministic Turing machines (DTM's) for small
 universal (deterministic)
Turing machines (U(D)TM's) was achieved via universal tag systems, needing exponential
space and time for the simulation.
 A polynomial amount of space and time was shown
in \cite{NW1,NW2}.

Universality has mainly been investigated intensively
for the class of Turing computability whereas there exist only few publications
on universality for restricted systems, having the same restriction.
Such a restriction is e.g. space or time complexity, or working conditions.
Furthermore,
the concept of universality also should be investigated for other underlying structures
than free monoids, e.g. commutative monoids or more general structures.
Actually the ideas presented here arose from the question if there exists a
universal (higher order) Petri nets capable of simulating all special Petri
nets but not having power of Turing computability.

The concept of universality discussed here is somehow restricted, since there
should not just be
a black box with input special machine (e.g. a DTM) together with its input,
 and output of the
somehow simulated result, without considering also e.g. space and time amount
of the simulation \cite{Min,DW}.
 It should also somehow simulate the (local) behaviour
(e.g. in space and time).

To clarify the problem universality for Turing machines will be considered in more
detail.

\section{General Turing Machines}

For general Turing machines (TM's), in particular for the deterministic
version (DTM), there exists the concept of {\it universality}. This
means that a single DTM can be constructed which can simulate
any special DTM in a certain way. For that purpose it is necessary
to encode a DTM as well as its input. Since the encodings have to be
invertible, the corresponding encoding functions have to be injective,
and the encoding as well as decoding functions have to be computable
by DTM's.

Formally, assume an enumerable universal alphabet $\Sigma$,
such that that any special DTM $M$ has a finite input and
work alphabet $\Sigma_M\subset\Sigma$. Similarly, assume an enumerable
universal set of states $Q$ with $Q\cap\Sigma=\emptyset$,
such that any special DTM $M$ has a finite
set of states $Q_M\subset Q$. In both cases, $\Sigma$ and $Q$ are
equivalent to ${\mathbb N}$.
It can be also assumed that $Q=\{q_1,\cdots ,q_{|Q_M|}\}$, and that
 the initial state of $M$ is $q_1$ and the
final state $q_{|Q_M|}$. Furthermore, that by renaming the symbols of
$M$ are $\Sigma_M=\{s_1,\cdots ,s_{|\Sigma_M|}\}$.

Any DTM $M$ defines a local function $f_M$ given by the program of $M$,
affecting the configurations of $M$.
Those can be written as $w_Lqxw_R$ where
$w_L,w_R\in\Sigma_M^*$, $x\in\Sigma_M$, and $q\in Q_M$.
An initial configuration has the form $c_1=w_Lq_1w_R$, and a final
 configuration the form $w_Lq_{|Q_M|}w_R$,
where $w_L\in\Sigma^*_M$ and $w_R\in\Sigma^+_M$. Quite often also
a normal form $c_1=q_1w$, and
$c_F=q_{|Q_M|}w$, respectively, is used where $w\in\Sigma_M^+$.
A step of $M$ can then be expressed by
$f_M(w_Lqxw_R)=w'_Lq'x'w'_R$,
with the usual interpretation of new symbol, new state and movement.
Thus
$f_M:\Sigma_M^*Q_M\Sigma_M^+\rightarrow\Sigma_M^*Q_M\Sigma_M^+$.
$f_M$ can be extended to a, in general partial, function
$f_M^*:\Sigma_M^*Q_M\Sigma_M^+\rightarrow \Sigma^*_MQ_M\Sigma_M^+$ by the definition

$f_M^*(w_Lq_1w_R)=w'_Lq_{|Q_M|}w'_R\ \Leftrightarrow\ \exists n\in {\mathbb N}:
f_M^n(w_Lq_1w_R)=w'_Lq_{|Q_M|}w'_R$.
This also defines a
function $\tilde{f}_M:\Sigma_M^+\rightarrow\Sigma_M^+$
by ($w=w_Lw_R$ and $w'=w'_Lw'_R$)
 $\tilde{f}_M(w)=w'\Leftrightarrow f_M^*(w_Lq_1w_R)=w'_Lq_{|Q_M|}w'_R$.
For the above mentioned normal form this gives
$f_M^*(q_1w)=q_{|Q_M|}w'\ \Leftrightarrow\ \exists n\in {\mathbb N}:
f_M^n(q_1w)=q_{|Q_M|}w'$
and

$\tilde{f}_M(w)=w'\Leftrightarrow f_M^*(q_1w)=q_{|Q_M|}w'$.

\medskip

For a universal DTM $U$ the program as well as the configurations of
special DTM $M$ have to be encoded. {\it Simple} encoding and decoding
functions should be used for that purpose. They have to be injective and computable
by DTM's $\Psi_M$ and $\Phi_M$, as well as the inverses by some
DTM's $\Psi'_M$ and $\Phi'_M$.
Actually, these $DTM$ are related to $M$ since some information on $Q_M$ and $\Sigma_M$
is necessary. Let $\Sigma_U\subset\Sigma$ be
the alphabet, and $Q_U\subset Q$ the set of states of $U$.
One possibility of encoding are functions
$f_{\Psi_M}:(\Sigma_M\cup Q_M)\rightarrow (\Sigma_U\setminus\{{\#}\})^+$ and
$f_{\Phi_M}:(\Sigma_M\cup Q_M)\rightarrow (\Sigma_U\setminus\{{\#}\})^+$
where ${\#}\in\Sigma_U$ is a special separator.

The initial and final configurations of $U$ for simulating $M$ can be chosen
in a normal form as
$q_1x\tilde{f}_{\Psi_M}(P_M){\#}y\tilde{f}_{\Phi_M}(c_1)$ and
$q_{|Q_U|}x\tilde{f}_{\Psi_M}(P_M){\#}y\tilde{f}_{\Phi_M}(c_F)$
where $x,y\in\Sigma_U$ are special symbols indicating the current position in
the program $P_M$ and configuration $c_M$, $P_M$ is the program of $M$,
and ${\#}\in\Sigma$ is a separator.
Both functions, $\tilde{f}_{\Psi_M}$ and $\tilde{f}_{\Phi_M}$ are injective.
The configurations of $U$ simulating $M$ have one of the forms

\noindent
$u_Lqu_Rxu'_R{\#}v_lyv_R$, $u_Lxu_Rqu'_R{\#}vLyv_R$,
$u_Lxu_R{\#}v_Lqv'_Lyv_R$, or

\noindent
$u_Lxu_R{\#}v_Lxv'_Lqv_R$, respectively.



For the simulation the following condition for one step
$c_M\rightarrow c'_M$ of $M$, or
reaching a final configuration from the the initial configuration of $M$ holds
(here for normal forms):

\noindent
$\forall M\forall c_M\exists n\in {\mathbb N}\
 :\ f^n_M(qxf_{\Psi_M}(P_M){\#}yf_{\Phi_M}(c_M))
 =q'xf_{\Psi_M}(P_M){\#}yf_{\Phi_M}(c'_M)$

or

\noindent
$\forall M\exists n\in {\mathbb N}\
 :\ f^n_U(q_1xf_{\Psi_M}(P_M){\#}yf_{\Phi_M}(c_1))
 =q_{|Q_U|}xf_{\Psi_M}(P_M){\#}yf_{\Phi_M}(c_F)$.

%
%

\bigskip

Often $\{0,1\}\subset \Sigma_U$ is chosen for the encoding alphabet.
$\Sigma_M$, $Q_M$, $P_M$, and configurations $c_M$ are encoded either
in unary or binary. Then $s_i$ can be encoded by $1^i$ in unary
or by $bin(i)$ in binary. Analogously for $q_j$.
A configuration $s_{i_1}\cdots s_{i_{\ell}}q_js_{i_{\ell+1}}\cdots s_{i_r}$ can be
 encoded e.g. by
 
$1^{i_1}0\cdots 1^{i_{\ell}}001^j01^{i_{\ell}+1}0\cdots 1^{i_r}0$ in unary or by

$bin(i_1)a\cdots bin(i_{\ell})c bin(j) c bin(i_{\ell+1})a\cdots bin(i_r)a$ in
binary where
$a,b$ are separating symbols. Encoding $0,1,a,c$ e.g. by $00,01,10,11$ then
gives a binary encoding using only $0,1$.

The unary encoding needs $|w|(|\Sigma_M|+1)+|Q_M|+4$ space, whereas the binary
encoding needs $2(|w|(\lceil log_2\Sigma_M\rceil +1)+\lceil log_2Q_M\rceil+2)$
space.

Encoding and also decoding can be achieved by very simple DTM's, actually by
deterministic finite state transducers (DFST's) consuming linear time.
For binary encoding the transducer also needs information on the sizes
$|\Sigma_M|$ and $|Q_M|$.


A special class of TM is the class of TM with alphabet $\{0,1\}$.
In that case only states have to be encoded. Thus the bounds
are $3|w|+|Q_M|+4$ or
$4|w|+2\lceil log_2|Q_M|\rceil +4$.

\bigskip

The considerations so far also hold for non-deterministic TM's (NTM's).
Encodings again can be achieved by DTM's. Instead of functions one has to
consider binary relations between configurations. The universal NTM
works like the deterministic counterpart, with the difference that because
of the encoding of the non-deterministic program $P_M$ the universal NTM
has choices for its steps.

It is also an interesting aspect to look for small universal NTM.

\section{Turing Machines with Constraint}

In the case of TM's with some constraint the DTM for encoding and
decoding should be restricted to the same constraint.
Such a constraint is e.g. space or time complexity. But it could be
also {\it reversibility} or some property of the function or relation
defined by the TM. In any case, besides alphabet and set of states
of the special TM also the complexity function has to be encoded.

In \cite{KM} it has been shown that for the class of DTM's with a binary
alphabet $\{0,1\}$ and a space complexity function belonging to a subclass
of primitive recursive functions over 1 variable, there exists a universal
DTM with the same complexity constraint. Note that only one variable is
needed for the computation of the complexity function.
Examples of such primitive recursive complexity functions are
$g(x)=c\cdot x$, $g(x)=c\cdot x^k$, $g(x)=c\cdot 2^x$, $g(x)=c\cdot 2^{2^x}$,
polynomials with non-negative coefficients
$$p(x)=\sum_{i=1}^kc_ix^{k-i}\ x_i\in {\mathbb N}\ ,$$
$g(x)=2^{p(x)}$ where $p(x)$ is a polynomial as just defined.

This subclass of primitive recursive functions $g$ does not use projections and
has the property:

$$\exists c_g\in {\mathbb N}\ \forall\bar{x}\in {\mathbb N}^k\ :
\ c_g+g(\bar{x})\geq\sum_{i=1}^kx_i\ ,$$

which in the case of 1 variable results in $c_g+g(x)\geq x$.

Thus $g(x)=\lceil log_2(x)\rceil$ does not belong to this subclass.

If $r_g$ is the length of the representation of the space complexity function
$g$ then the universal DTM needs only $s_g(x)=g(x)+r_g$ space.

For the analogous time complexity constraint there is a price of polynomial
cost. In this case there is a polynomial $p_g$ with non-negative coefficients
for every primitive recursive
function from the subclass defined above such that the time complexity
of the universal DTM fulfills $t_g(x)\leq p_g(g(x))$ for all $x\in {\mathbb N}$.

Since the space and time complexity functions are computed by DTM's it follows
by straight forward arguments that the results from above also hold for
non-deterministic TM's (NTM's), i.e. there exists a universal NTM for the subclass
of primitive recursive functions as complexity constraints.

Thus there exist universal DTM's and NTM's with complexity constraints for
each complexity function from the special class of primitive recursive
functions (for time complexity only with additional polynomial).
In particular, there exist universal (N)LBA's and DLBA's.

\section{Finite Automata and Finite State Transducers}

In the case of deterministic and non-deterministic finite automata (DFA's, NFA's),
as well as deterministic and non-deterministic finite state transducers (DFST's, NFST's)
the encoding and decoding of special FA and FST should be achieved by DFST's.
Otherwise, the encoding might contain too much computation power, going 
even beyond the power of FA's or FST's.

In \cite{Kud} it was shown that under the above condition that encoding
of special FA's and FST's has to be achieved by a DSFT, there exists no universal DFA,
NFA, DFST and NFST. The proof uses a contradiction on the number of states of such
a universal automaton.

Thus there are no universal FA's in this strong sense.

In the same paper (\cite{Kud}) it has also been shown that there exists a universal FA for
the class of all FA's with a bounded number $k$ of states. However, that universal
FA does not belong to the same class, needing strictly more states.

\section{Pushdown Automata and Pushdown Transducers}

For universal DPDA's and NPDA's, as well as for corresponding transducers (DPDT's and NPDT's),
the encoding of special DPDA and NPDA should be done
by a DPDT. From the considerations above for Turing machines
 it follows that deterministic finite state
transducers suffice for encoding and decoding.
It is unknown that under such conditions there exist universal automata
of that kind. But it is my conjecture that such do not exist, since there is a problem in
accessing the program of a special PDA.
Possibly this can be shown using Kolmogorov complexity arguments.

\section{Rewriting Systems}

In the case of word grammars or other word rewriting systems like L-systems,
it is necessary to use a deterministic or monogeneous grammar or rewriting system
of the same type for encoding the special ones.

For general semi Thue systems (STS) or type-0 grammars this is straight forward since
any such system can be simulated by a TM, and vice versa.
In a similar way, type-1 or monotone frammars can be simulated by NLBA's and
vice versa.

For type-2 grammars, however, my conjecture is that there does not exist a universal
grammar of that type. Here the encoding should be achieved by some kind of deterministic
or monogeneous type-2 grammar used as transducer.
The same seems to hold for ETOL systems where encoding should be done by an EDTOL system
used as transducer.
An ETOL system \cite{RS} consists of a finite number of not necessarily
disjoint finite sets of context-independent rules which in one derivation step
have to be applied in parallel to all symbols in a word using only, not necessarily
identical rules, of one set. In an EDTOL system there is only one rule for each
symbol.

\section{Multiset Grammars}

Whereas for word grammars the encoding of special grammars can be done by a FST
it is a problem for multiset grammars, at least for simple multisets as elements from
${\mathbb N}^k$. This arises when looking for a universal vector addition system (VAS)
or a universal Petri net (PN). Most probably, such don't exist, at least for simple
multisets, i.e. elements from ${\mathbb N}^k$. The reason for that possibly is the
commutativity of the basic underlying operation $\oplus$, whereas for word systems catenation
is not commutative.

Possibly, higher order multisets, used as tokens in higher order Petri nets,
might be a solution for the construction of universal higher order Petri nets.
However, they should not be of the power of Turing machines. 

\section{General Universality}

A possible definition of general universality to cover e.g. all the cases
considered so far, may be the following construction.

Let ${\cal M}$ be a universe of elements, equipped with a binary operation $\circ$
  and ${\cal R}$ a collection (class)
of binary relations
 $r\subseteq {\cal M}_r\times {\cal M}_r$ with the property
 $\circ:{\cal M}_r\times {\cal M}_r\rightarrow {\cal M}_r$ for $r\in {\cal R}$.

To define a universal relation $R$ for ${\cal R}$ encoding of elements
$x\in {\cal M}_r$ and of $r$ itself is necessary.
This might be achieved by injective functions
$f_r:{\cal M}_r\rightarrow {\cal M}'_R\subset {\cal M}$
 and $g:{\cal R}\rightarrow {\cal M}''_R$ 
from the same class ${\cal R}$,
such that

$\forall r\in {\cal R}\forall x,y\in {\cal M}_r\exists n\in {\mathbb N}\ :
(g(r)\circ f_r(x),g(r)\circ f_r(y))\in R^n\
\Leftrightarrow\ (x,y)\in r$,
where ${\cal M}_R={\cal M}'_R\cup {\cal M}'_R$ and
 ${\cal M}'_R\cap {\cal M}''_R=\emptyset$, and
 $R\subseteq {\cal M}_R\times {\cal M}_R$.

But there is a problem about the operation $\circ$ since actually it is
a ternary relation on ${\cal M}$.

More general, with a single pairing function
 $\phi:{\cal M}\times {\cal M}\rightarrow {\cal M}_R$,
 the condition is
 
$\forall r\in {\cal R}\forall x,y\in {\cal M}_r\exists n\in {\mathbb N}\ :
\ (g(r),\phi(f_r(x),f_r(y))\in R^n\ \Leftrightarrow\ (x,y)\in r$.

Another possibility may be to choose as basic universe a Scott domain as
for the construction of a model for the $\lambda$-calculus.

\bibliographystyle{eptcs}

\begin{thebibliography}{}
\providecommand{\bibitemstart}[1]{\bibitem{#1}}
\providecommand{\bibitemend}{}
\providecommand{\bibliographystart}{}
\providecommand{\bibliographyend}{}
\providecommand{\url}[1]{\texttt{#1}}
\providecommand{\urlprefix}{Available at }
\providecommand{\bibinfo}[2]{#2}
\bibliographystart

\bibliographyend
\end{thebibliography}


\begin{thebibliography}{99}

\bibitem{DW}Davis, M. D., Weyuker E. J.: {\it Computability, Complexity, and
Languages: Fundamentals of Theoretical Computer Science.} Academic Press, 1983.

\bibitem{KM}Kudlek M., Margenstern, M.: {\it Universal Turing Machines with
Complexity Constraints.} Proc. Intern. Conf. {\it Automata and Formal Languages
VIII}, Publ. Math. Debrecen {\bf 53}, pp. 895-904, 1999.

\bibitem{Kud}Kudlek, M.: {\it On Universal Finite Automata and a-Transducers.}
(In: {\it Grammars and Automata for String Processing: from Mathematics and Computer
Science to Biology and back.} Eds. C. Mart\'{\i}n Vide, V. Mitrana. {\it Topics
in Computer Mathematics}, pp. 163-170, Taylor and Francis, London, 2003.)

\bibitem{Min}Minsky, M.L.: {\it Computation: Finite and Infinite Machines.}
Prentice Hall, 1972.

\bibitem{NW3}Neary, T., Woods, D.: {\it The P-completeness of Cellular Automton
Rule 110.} Proc. ICALP 2006, LNCS {\bf 4051 I}, pp.132-143, 2006.

\bibitem{NW4}Neary, T., Woods, D.: {\it On the Time Complexity of 2-tag Systems and
Small Universal Turing Machines.} Proc. FOCS 2006, pp. 439-446, 2006.

\bibitem{NW1}Neary, T., Woods, D.: {\it Small Fast Universal Turing Machines.}
TCS {\bf 362(1-3)}, pp. 171-195, 2006.

\bibitem{NW2}Neary, T., Woods, D.: {\it Four Small Universal Turing Machines.}
Proc. MCU 2007, LNCS {\bf 4664}, pp. 242-254, 2007.

\bibitem{RS}Rozenberg, G., Salomaa, A. K.: {\it The Mathematical Theory of L systems.}
Academic Press, 1980.

\end{thebibliography}

\end{document}